# Threat Intelligence Driven IP Protection for Entrepreneurial SMEs

*Short Paper*

**Sam Pitruzzello**
University of Melbourne
Australia
s.pitruzzello@student.unimelb.edu.au

**Atif Ahmad**
University of Melbourne
Australia
atif@unimelb.edu.au

**Sean Maynard**
University of Melbourne
Australia
seanbm@unimelb.edu.au

## Abstract

*Entrepreneurial small-to-medium enterprises (E-SMEs) face significant cybersecurity challenges when developing valuable intellectual property (IP). This paper addresses the critical gap in research on how E-SMEs can protect their IP assets from cybersecurity threats through effective threat intelligence and IP protection activities. Drawing on Dynamic Capabilities and Knowledge-Based View theoretical frameworks, we propose the Threat Intelligence-driven IP Protection (TI-IPP) model. This conceptual model features to modes of operation — closed IP development and open innovation – enabling E-SMEs to adapt their IP protection and knowledge management strategies. The model incorporates four key phases: sensing opportunities and threats, seizing opportunities, knowledge transfer, and organizational transformation. By integrating cybersecurity threat intelligence with IP protection practices, E-SMEs can develop capabilities to safeguard valuable IP while maintaining competitive advantage. This research-in-progress paper outlines a qualitative research methodology using multiple case studies to validate and refine the proposed model for practical application in resource-constrained entrepreneurial environments.*

## Introduction

Entrepreneurial small-to-medium enterprises (E-SMEs) play an important role in national economies contributing to economic growth, innovation and employment. Accounting for over 90% of businesses in many countries, their success is crucial from a social and economic perspective (Arroyabe et al., 2024). More impressively, high growth E-SMEs, which comprise a small percentage of all firms, deliver outsized economic contributions and performance (Raby et al., 2022). To drive growth, E-SMEs need to develop and commercialize intellectual property (IP) and create innovative products, services and technologies (Chesbrough, 2003). We define E-SMEs as companies that experience high levels of growth, have the potential to scale globally and develop technological innovations.

During periods of rapid growth, E-SME's are exposed to cybersecurity threats (Raby et al., 2022) and experience higher failure rates compared to larger firms (Shane, 2009). Maintaining a focus on growth and commercial viability often overshadows cybersecurity concerns despite the risks (Chidukwani et al., 2024; Olander et al., 2011). These problems are compounded by several factors unique to E-SMEs - limited financial resources, lack of dedicated cybersecurity personnel, over-reliance on IT outsourcing (Bada & Nurse, 2019; Osborn et al., 2018) and striking the right balance between engaging in open innovation and protecting IP assets (Grimaldi et al., 2021). These factors make E-SMEs particularly vulnerable to cybersecurity risks from a variety of threat actors, including competitors, organized crime syndicates and nation-states who understand the value of IP (Kotsias et al., 2022). Recovering from common cybersecurity incidents is becoming easier provided effective technical systems and operational practices are in place such as regular operating system patching, malware, strong password policies and backups (Chandna & Tiwari, 2021). However, loss or theft of valuable IP can have a disastrous effect on an E-SMEs business potentially leading to failure (Chidukwani et al., 2024). The economic impact of IP theft and cyber-espionage is significant with the IP Commission report stating that "IP theft costs the U.S. economy hundreds of billions of dollars annually and reduces U.S. companies' R&D investment and innovation" (Blair et al. 2021, p. 1). In Europe, theft of IP and counterfeit products was estimated at EUR338 billion in 2013 (Europol, 2013).

Cybersecurity research has focused on large enterprise despite the challenges faced by SMEs (Armenia et al., 2021; Arroyabe et al., 2024). There is also limited research examining how E-SMEs can protect their IP from cybersecurity threats. Finally, there is little research on the intersection of threat intelligence and IP protection (IPP). We argue that E-SMEs need to base strategies on real-world, actionable information. Therefore, this paper addresses these gaps by discussing how E-SMEs can defend against IP theft from advanced cybersecurity threats. We draw on two theoretical lenses, Dynamic Capabilities (DC) and Knowledge-Based View of the firm (KBV) to propose a conceptual model that considers threat intelligence as a key driver in developing IPP capabilities. For the preceding reasons, the research question is:

> *How can entrepreneurial SMEs leverage threat intelligence to protect their valuable IP from cybersecurity threats?*

The remainder of this research-in-progress paper is structured as follows. First, an overview of the research methodology used in the literature review is covered. This is then followed by a discussion on the theoretical background and literature review. The conceptual model is then presented before discussing the conclusion and future research for the project.

## Research Methodology

A literature review drawing from Okoli's (2015) systematic literature review methodology was conducted (Okoli, 2015). The search and review provide an assessment of the latest research grouped under several core concepts and theoretical frameworks. The process of how articles were found, including the search terms, and analyzed is described in this section. The first step involved a Google Scholar search, resulting in 5,509 articles. The search was conducted over a three-month period between late January and early April 2025. A combination of the following search terms was used in the search:

> ("cybersecurity" OR "cyber-attack" OR "cybersecurity incident"); ("small to medium enterprise" OR SME OR "small to medium business" OR "startup" OR "start-up"); (entrepreneur* OR "entrepreneurial" OR "entrepreneurship"); ("Intellectual Property Protection" OR "IP Protection"); ("cybersecurity threat intelligence")

Google Scholar search ranks articles by relevance so the first 10 pages (100 articles) were selected in the first filter step. Articles were then filtered by excluding theses, books and non-peer reviewed articles resulting in 68 articles (filter step 2). The list was extended to include high-quality articles in a) the reference list of short-listed papers and b) seminal works in entrepreneurship, innovation and cybersecurity. An additional 28 papers were included for a total of 96 papers. These papers were filtered by applying a third screen according to the following inclusion criteria (filter step 3) - articles focusing on **cybersecurity in SMEs, IP protection in SMEs, cybersecurity threat intelligence and the application of organizational theories to cybersecurity**. The final result was a list of 36 papers that were analyzed for inclusion in the literature review. Table 1 below outlines the search process and results. From these 36 articles, 19 have been used in the literature review (Background section) and conceptual model design while the remaining 17 papers provide evidence and support in other areas of the paper.

| Step | Filter | No. of papers |
|------|--------|---------------|
| Filter Step 1 | Top search results from Google – most relevant/top ranked papers | 100 |
| Filter Step 2 | Remove books, theses and non-peer reviewed articles | 68 |
| | Additional papers selected from reference list of 68 short-listed papers | 28 |
| | Total papers short listed for detailed review | 96 |
| Filter Step 3 | **Short-Listed papers after filtering by final selection criteria** | **36** |
| | **Table 1 - Literature search and filtering results** | |

# Background

Relevant areas of literature are discussed in this section. First, and from a theoretical perspective, Dynamic Capabilities (DC) and Knowledge-based View (KBV) are covered. Cybersecurity threat intelligence (CTI) is then discussed followed by IPP in E-SMEs. This provides the basis for developing a conceptual model. DC (Teece et al., 1997) and KBV (Grant, 1996) have been chosen as the theoretical lenses for several reasons. First, they cover key characteristics of knowledge that are pertinent for organizations to create value. Secondly, they are simple concepts and easy to apply in practice. Finally, these theories have been used extensively in organization research particularly in the fields of entrepreneurship, innovation (Grant, 1996; O'Reilly III & Tushman, 2013; Teece et al., 1997) and cybersecurity (Ahmad et al., 2014; Naseer et al., 2024; Pigola & Rezende Da Costa, 2023).

## *Dynamic Capabilities & Knowledge-Based View of the Firm*

DC states that firms need to change and adapt their resources in line with changing market and economic conditions. The ability to reconfigure assets to maintain a competitive advantage is considered a dynamic capability (Teece et al., 1997). DC builds on the Resource-Based View of the firm, considered a 'static' theory (Hernández-Linares et al., 2021) and allows organizations the "…ability to integrate, build and reconfigure internal and external competences to address rapidly changing environments" (Teece et al., 1997, p. 516). Teece (2007) enhanced DC by introducing two additional microfoundations to support resource and organizational transformation – sensing opportunities and threats and seizing opportunities (Teece, 2007). DC can be applied to the dynamic nature of managing and protecting IP during rapid growth.

KBV states that knowledge, both explicit and tacit, is held by individuals and leveraged by firms for productive output and to gain a competitive advantage. Explicit knowledge, also referred to as codified knowledge, is knowledge that can easily be recorded and stored in information Systems (IS) in the form of documents, plans, designs, procedures and processes. Tacit knowledge is held in employees' heads and difficult to store, publish and share since it is a person's 'know-how' and forms the basis of their skills (Grant, 1996; Olander et al., 2011). Tacit knowledge is shared through demonstration, practice and teaching and requires feedback. KBV involves three key processes – knowledge absorption, knowledge transfer and knowledge appropriation. Knowledge absorption is the capacity for an individual or organization to take in and synthesize knowledge. Knowledge transfer is the dissemination and communication of knowledge. Finally, knowledge appropriation is the process of creating value from knowledge that has been absorbed and transferred (Grant, 1996). DC and KBV can be applied to the dynamic environment in which E-SMEs operate. In the following sections, cybersecurity threat intelligence (CTI) and IPP in E-SMEs are discussed.

## *Cybersecurity Threat Intelligence*

CTI is the collection, analysis, and dissemination of information about current and potential cybersecurity threats that can impact an organization's assets and operations. It involves gathering data on threat actors, their tactics, techniques, and procedures, and using this information to make informed decisions that strengthen the organization's security posture. NIST (National Institute of Standards and Technology) Special Publication 800-150: Guide to Cyber Threat Information Sharing defines threat intelligence as "… threat information that has been aggregated, transformed, analyzed, interpreted, or enriched to provide the necessary context for decision-making processes." (Johnson et al., 2016, p. 2). CTI helps organizations develop a proactive and targeted approach to safeguarding digital assets and operations. Key strategies for effective CTI include developing situation awareness, utilizing Artificial Intelligence (AI) and incorporating cybersecurity knowledge into threat intelligence processes (Arikkat et al., 2024). Situation Awareness is critical for developing practices that enhance E-SMEs awareness of both the cyber-threat landscape and the broader business context (Ahmad et al., 2021). E-SMEs can leverage AI-driven solutions to identify and extract information from various sources to generate organization-specific threat intelligence knowledge.

To enhance understanding of cybersecurity threats, E-SMEs need to consider approaches that incorporate cybersecurity knowledge into their threat intelligence processes. This can involve building a cybersecurity knowledge base and training cybersecurity-aware AI knowledge agents (Wang et al., 2024). We argue that CTI is a knowledge management function and by implementing CTI, E-SMEs can significantly improve their ability to protect valuable IP from cybersecurity threats. The key is to develop a tailored, knowledge-driven approach to cyber-threat intelligence that aligns with the E-SME's specific needs.

### Protecting IP in E-SMEs

IP can be defined as "the creation, ownership and control of original ideas as well as the representation of those ideas" (Whitman & Mattord, 2014, p. 52). IP assets, like any organizational asset, enable E-SMEs to derive value by commercializing innovative products and services or licensing them to external firms (Amara et al., 2008; Chesbrough, 2003; Grant, 1996). Therefore, IPP is an important activity in any entrepreneurial venture. IPP measures (IPPM) can be formal or informal. Formal methods are backed by laws and include trademarks, patents, copyright and registered designs. They provide the IP owner rights to generate exclusive revenues from products and services derived from their IP. If a competitor infringes on another firm's IP by copying, counterfeiting or reverse engineering, the IP owner can sue for damages and loss of revenue. Informal methods do not have legislative protections, so they employ organizational capabilities and strategies such as design complexity, trade secrets and lead-time advantage to protect their IP (Amara et al., 2008; Olander et al., 2011). The challenge for E-SMEs is determining appropriate IPPM and associated cybersecurity practices that enhance IPPM. Cybersecurity practices can protect IP assets from sophisticated threat actors and advanced persistent threats (APT) from accessing digital IP assets (Ahmad et al., 2019). A conceptual model is proposed, designed with the needs of E-SMEs in mind, driven by information gathered during knowledge-based threat intelligence activities.

# Conceptual Model

Figure 1 on the following page illustrates the conceptual model – Threat Intelligence-driven IP Protection (TI-IPP). Teece (2007) suggests that organizational assets need to be transformed to maintain competitive advantage by sensing and seizing opportunities (Teece, 2007). TI-IPP integrates the three core concepts from DC and one from KBV – sensing opportunities and threats (DC), seizing opportunities (DC), transforming the organization (DC), and knowledge transfer (KBV) (Grant, 1996). Comparing DC and KBV, sensing and seizing (DC) and knowledge absorption (KBV) are similar concepts. Transform (DC) and knowledge appropriation (KBV) are also comparable. Knowledge transfer (KBV) has been added in the model since it is unique and describes how knowledge is shared and communicated within and external to the organization. The reason for expanding DC to include knowledge transfer is to introduce a key feature of the model where it can be switched from 'closed IP development mode' to 'open innovation mode'.

From an IS perspective, extending DC to include KBV's knowledge transfer activity enhances knowledge protection from a cybersecurity point of view. DC and KBV have been extensively applied in IS research and DC has only recently been applied to cybersecurity research (Naseer et al., 2024). Furthermore, knowledge management focuses on open sharing of knowledge with a limited view of protecting valuable organizational knowledge (Ahmad, et al., 2014). Drawing on DC and KBV and applying stakeholder-activity matrix model (Ahmad et al., 2011), four roles are considered in the model - governance, leadership, security and knowledge management. Each activity in the model has been aligned to each role. One of the outcomes of the research is to validate these assumptions.

The model begins in closed IP development mode as this is the natural initial state for E-SMEs – the creation of new ideas and concepts. During this phase, E-SMEs haven't developed IP sufficiently enough to warrant formal IPPMs. Therefore, informal IPPMs such as non-disclosure agreements (NDAs), Memorandum of Understanding (MOU) and trade secrets are more suitable during the early stages of IP development (Grimaldi, et al., 2021; Toma, et al., 2018). As the innovation process progresses, formal IPPMs can be implemented, and the E-SME can engage in open innovation to commercialize IP. Grimaldi et al. (2021) and Toma et al. (2018) provide insights into how the model can be switched from closed to open modes – by considering when the E-SMEs is ready to commercialize its IP. Starting at Sense, governance gathers competitive intelligence while leadership assesses the IP asset inventory. Leadership, in conjunction with security, conduct threat and vulnerability assessments. Security works with knowledge management to collate knowledge on cyber security threats and internal knowledge on IS vulnerabilities.

Next comes Seize, where governance is responsible for approving IP asset acquisition while leadership will carry out new IP asset acquisition or development. The security role will review IP assets to determine adequate IPPM and knowledge management ensures that IP assets are adequately stored and accessible. Transfer consists of three key activities in 'closed IP Development' mode – develop controlled disclosure protocols (driven by governance and leadership), sanitize knowledge (leadership and security role) and implementing secure communications and access (security roles).

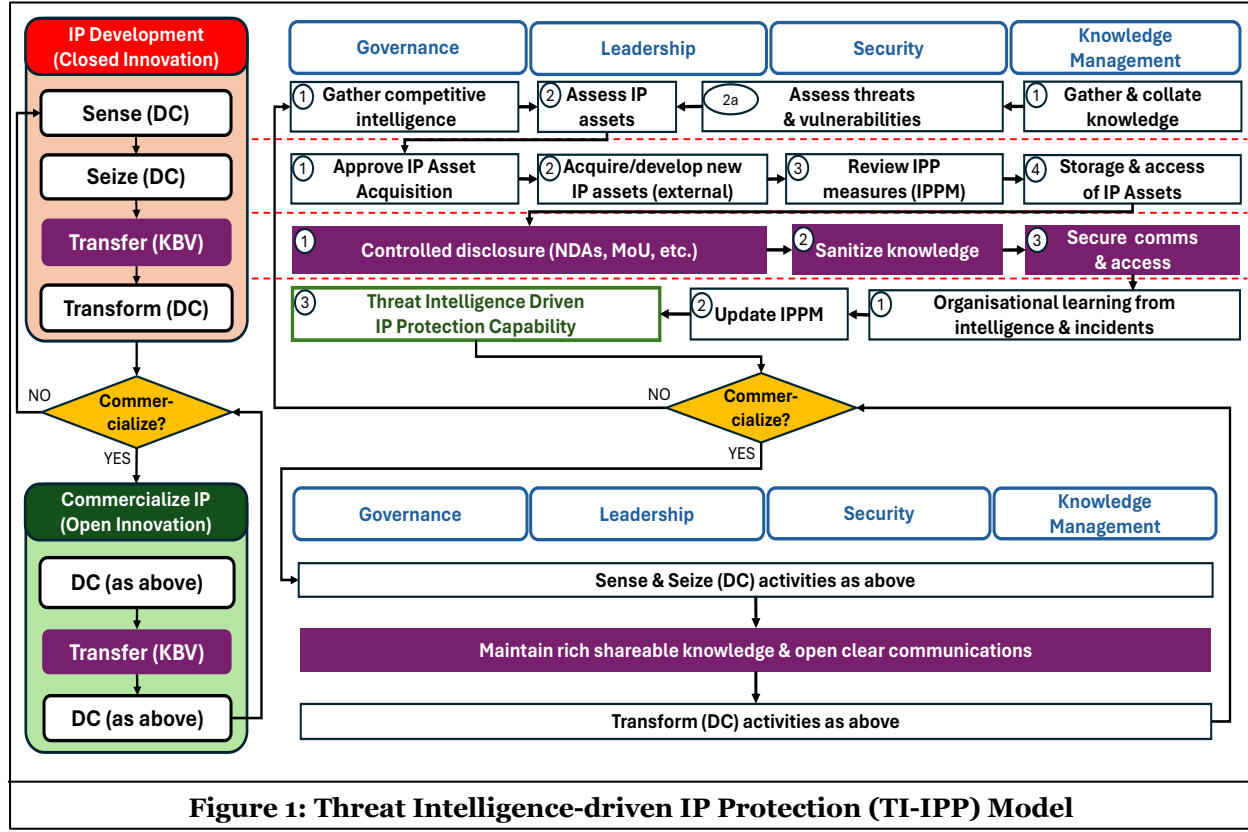

**Figure 1: Threat Intelligence-driven IP Protection (TI-IPP) Model**

In 'open innovation' mode, the Transfer phase has one activity – maintain risk shareable knowledge and open communications. Finally, in the Transform row knowledge management and security roles are responsible for organizational learning while leadership and security update IPPM. The final activity is the capability developed by the E-SME as a result of the previous activities – threat intelligence-driven IPP capability. The DC activities are repeated in 'open mode' and shown simplified in the bottom part of the model. Each activity is discussed in further detail in the following sections with Table 2 below summarizing each phase of the model, the associated activities and references from the literature.

| Model Phase | Activity | References |
|---|---|---|
| Sense | Competitive Intelligence | Ahmad et al., (2019); Ahmad et al. (2021); Calof & Wright (2008); Crane (2005); Grimaldi et al. (2021); Teece (2007) |
| | Assess IP Assets | |
| | Assess Threats & Vulnerabilities | |
| | Gather & Collate Knowledge | |
| Seize | Approve IP Asset Acquisition | Chesbrough (2003); Grimaldi et al. (2021); Teece (2007) |
| | Acquire or Develop New IP Assets | |
| | Review IP Asset Protection Measures | |
| | Store & Access IP Assets | |
| Transfer | OPEN: Maintain rich shareable knowledge | Decision/switch from open to closed: Grimaldi et al. (2021); Toma et al. (2018) |
| | CLOSED: Controlled Disclosure Protocols | |
| | CLOSED: Sanitize Knowledge | |

| | CLOSED: Secure Communications & Access | Ahmad et al. (2014); Grant (1996) |
|---|---|---|
| Transform | Organizational Learning | Ahmad et al. (2014); Teece (2007) |
| | Update IP Protection Measures (IPPM) | |
| | Threat Intelligence-driven IPP Capability | |
| **Table 2 – Phase and activities in the research model** | | |

### Sensing Opportunities and Threats (DC)

Sensing opportunities and threats is one of the micro-foundations of DC and the first phase in developing new capabilities to maintain competitive advantage (Teece, 2007). From a cybersecurity perspective, E-SMEs need to be aware of threat actors looking to acquire their valuable IP assets (Ahmad et al., 2019). This phase in the model involves four activities – obtain competitive intelligence, gather and collate organizational knowledge, assess threats and vulnerabilities and assess IP asset inventory. To compete in global markets, E-SMEs need to obtain competitive intelligence (CI) on their competitors while denying them access to their knowledge and IP (Calof & Wright, 2008). E-SMEs also need to assess the risks of threat actors looking to acquire their IP assets or infringing on their IP rights. Assessing threats and vulnerabilities requires gathering a large amount of information both externally and from internal IS, analyzing the information and making sense of it (Ahmad et al., 2021). In addition, it is crucial that E-SMEs know where and how IP assets are stored and who has access. E-SMEs need to ascertain whether their portfolio of IP assets is adequate in light of their competitive intelligence discoveries. If gaps exist in their IP asset portfolio, they need to be developed or acquired (Grimaldi et al., 2021).

### Seize Opportunities (DC)

Seizing opportunities in DC recommends that organizations need to seize new opportunities once they have been sensed and identified. This phase of the model involves four activities – approving IP asset acquisition and/or development, acquiring or developing new IP assets, reviewing IPPM and configuring IS for adequate storage and access of IP assets. The development and commercialization of IP requires considerable time, effort, resources and investment (Chesbrough, 2003; Teece, 2007). Similarly, implementing sound IPPM requires significant resources and expertise. The resources and expertise required in successfully executing these activities can be a limiting factor for many E-SMEs. A more expedient way for E-SMEs to implement these initiatives is to partner with external organizations and experts. Where existing relationships don't exist, the E-SME will need to establish new partnerships to assist with commercialization activities (Chesbrough, 2003) and engage with external experts and professional service providers to implement IPPM (Grimaldi et al., 2021). Finally, the storage and accessibility of IP assets is implemented within the organization's IS.

### Transfer (KBV)

Efficient knowledge transfer within an E-SME is critical to gaining a competitive advantage (Grant, 1996). However, E-SMEs need to be diligent not to share or allow access to sensitive knowledge and IP during its development. Therefore, we propose that this phase of the model be activated in one of two modes – open innovation and closed IP development mode. In closed IP development mode, knowledge needs to be securely transferred within the organization and tightly controlled externally. The three activities in this mode are implementing disclosure protocols, sanitizing knowledge and implementing secure communications and access. Implementing information disclosure protocols involves classification of documents, restricting access to data and information on need-to-know basis and drafting contracts such as non-disclosure agreements (NDAs) and non-compete clauses in commercial and employment contracts (Ahmad et al., 2014). Sanitizing knowledge involves several important activities including but not limited to converting tacit knowledge to codified knowledge for storage and sharing, anonymizing personally identifiable information and encrypting sensitive information. Secure communications and access to IS involves encrypting information for secure storage and communication and implementing access control policies including role/rule-based access and zero-trust policies to control and monitor access. In the open innovation mode, the organization switches to traditional knowledge management function of maintaining and sharing knowledge using clear, open and easy to access IS (Ahmad et al., 2014).

### Transform (DC)

The final phase in the model involves the transformation of the E-SME into one that can build a capability to protect IP based on threat intelligence activities. Once opportunities and threats have been sensed and seized, knowledge transferred throughout the organization and IP developed in a secure manner, the E-SMEs can transform their cybersecurity capabilities to protect their valuable IP (Ahmad et al., 2014; Teece, 2007). E-SMEs with threat intelligence-driven IPP capability will be able to develop and commercialize IP securely. There are two activities in this phase of the model - capturing organizational learning from intelligence activities and updating IPPM. Capturing learning from intelligence activities involves both threat intelligence and competitive intelligence gathering activities. Updating IPPM is an ongoing process of reviewing and assessing the IPPM's in place in light of new knowledge obtained from intelligence gathering activities.

## Conclusion and Future Research

During periods of high growth, E-SMEs experience rapid change, high levels of uncertainty and the need to navigate tough business environments with limited resources. Furthermore, they constantly face competing priorities and need to strike a balance between engaging in open innovation and protecting their valuable IP. We propose a conceptual model that shows how E-SMEs can transform their cybersecurity practices to protect IP by considering threat intelligence. Drawing on DC and KBV, four key phases were designed into the model – sensing opportunities and threats, seizing opportunities, knowledge transfer and transform. The aim of this project is to validate the model to ensure suitability for E-SMEs with the potential to expand its use to any type of SME that needs to protect commercially sensitive information.

This research will address a critical gap, since there has been insufficient research that addresses the cybersecurity needs of smaller entrepreneurial organizations. From a theoretical perspective, DC has been extended to include knowledge transfer from KBV for a key reason – it will inform the 'mode switch' where an E-SME locks down communications and access to sensitive information during the IP development stage then switches to open innovation model when commercializing IP. This research project is primarily focused on developing a model that can be applied in practice to SMEs. In addition, theoretical contributions will be made to DC by determining how it may be extended to incorporate elements of KBV and the application of these theories to cybersecurity and IP protection.

### Future Research

This research is based on qualitative research methods with multiple case studies. Eight to twelve case studies will be used to answer the research question and validate the conceptual model (Yin, 2017). The activities in the conceptual model were derived from the literature with a focus on competitive intelligence and CTI activities of which many E-SMEs undertake in their normal operations. The main goal for this research project is to develop a model that will be easy for E-SMEs to implement by minimizing the use of overly complex standards and frameworks typically applied in large organizations. Up to four representatives from each case will be interviewed using semi-structured interviews. Interview questions will focus on the following areas - how the organization collects intelligence on cybersecurity threats, the measures in place to ensure IP is protected and how the organization ensures appropriate access and disclosure of commercially sensitive information. The individuals in case study organizations will be senior leadership executives, founders and board directors. Collectively, the individuals will have thorough insight and knowledge across the entire organization. While the focus of the study is on senior leaders and founders due to their in-depth knowledge and history on how the organization has developed over time, the study will not exclude interviews with junior employees if they can add value to the study. Focus group workshops will be held to validate the improved conceptual model with representatives from each of case study invited to participate. Case study participants will be selected according to the following criteria:

1. The case unit is the **organization** (E-SMEs)
2. Case units share similar structure (**size = SME**) and located in Australia (**location: AU**)
3. Case study participants develop valuable IP and innovative technologies

In terms of analyzing data, this project adopts Yin's (2017) approach of first designing a case study interview protocol to answer the main research question and validate each activity in the model. Yin (2017) details four general strategies (theory based, case descriptions, examining rival explanations and ground up) and five analytic techniques (pattern matching, explanation building, time-series analysis, logical models and cross-case synthesis) to analyze qualitative data. The analysis strategy best suited for this research is theory-based given that the model has been derived from two theoretical frameworks. Two analysis techniques are suitable for this research project - logical models and cross-case synthesis. Logical models operationalize complex chain of events and cross-case synthesis techniques are applicable to multiple case studies. Since this research project aims to validate a model with a sequence of activities by investigating a number of case studies, these two techniques are suitable.

The main limitation of this study is the single-country focus, that could limit generalizability. However, some of the case study organizations that have agreed to participate have expanded their operations into overseas markets. Therefore, this may present an opportunity to minimize single country bias. Two pilot case studies have been completed with a view to completing data collection by October this year. Analysis and preliminary findings are expected to be completed by the end of the year. Ethics approval has been received from the research institution covering data collection and the interview protocol.